\documentclass{appolb}
\usepackage{epsfig}

\begin{document}
\title{Taylor's power law for fluctuation scaling in traffic
\thanks{Presented at the Summer Solstice 2009 International Conference
on Discrete Models of Complex Systems Gda\'nsk, Poland, June 22-24, 2009.}}

\author{Agata Fronczak, Piotr Fronczak, Maksymilian Bujok
\address{Faculty of Physics, Warsaw University of Technology,\\
ul. Koszykowa 75, 00-662 Warsaw, Poland}}

\maketitle

\begin{abstract}
In this article, we study transportation network in Minnesota. We show that the system is characterized by Taylor's power law for fluctuation scaling with nontrivial values of the scaling exponent. We also show that the characteristic exponent does not unequivocally characterize a given road network, as it may differ within the same network if one takes into account location of observation points, season, period of day, or traffic intensity. The results are set against Taylor's fluctuation scaling in the Nagel-Schreckenberg cellular automaton model for traffic. It is shown that Taylor's law may serve, beside the fundamental diagram, as an indicator of different traffic phases (free flow, traffic jam etc.).
\end{abstract}

\section{Introduction}\label{intro}

In ecology, a striking observation is that variability in population abundance of a species and average population density are related in both space and time. The relationship is known as the Taylor's power law (or the law of the mean) and states that the mean $\langle N\rangle$ and the variance $\langle N^2\rangle-\langle N\rangle^2$ characterizing the number of population representatives
are related by the power law
\begin{equation}\label{taylor}
\langle N^2\rangle-\langle N\rangle^2=a\langle N\rangle^b,
\end{equation}
with the characteristic exponent $b$ describing effects of heterogeneity in spatial or temporal
patterns of the frequency distribution.

Since the publication of Taylor's famous paper in 1961 \cite{1961NatureTaylor}, the law of the mean has been substantiated in a vast body of ecological data that spans from protozooans to human populations \cite{1978JAnimTaylor}. This kind of fluctuation scaling has been also noticed in a variety of natural phenomena including precipitation in a given area and river flows \cite{2008AdvEisler}. Taylor's law has also been observed in dynamics of different man-made systems driven by human activity \cite{2004PRLBarabasi}. Here, the list of documented examples includes behavior of the Internet and stock market dynamics.

In this paper we perform a detailed analysis of fluctuation scaling in car traffic. We show that fluctuations in the number of cars passing through a given observation point fulfil the relationship given by Eq.~(\ref{taylor}). The clue observation reported in this paper is that the scaling exponent $b$ does not unequivocally characterize a given road network. The parameter $b$ may differ not only among various transportation networks but also within the same network if one takes into account: location of observation points (intersection, one-way street, highway, etc.), season, period of day, or simply traffic intensity. For this reason Taylor's law may serve, beside the fundamental diagram, as an indicator of different traffic phases (free flow, traffic jam etc.).

The outline of the paper is as follows: At the beginning (Section~\ref{SecReal}) we analyze Taylor's fluctuation scaling in real data. In Section~\ref{SecNaSch} we show that Taylor's law is also observed in the simplest cellular automaton model for traffic (the Nagel-Schreckenberg model), although interpretation of the obtained results deserves further studies. Section~\ref{SecSum} gives concluding remarks.

\section{Fluctuation scaling in real traffic}\label{SecReal}

\begin{figure}
\centerline{\epsfig{file=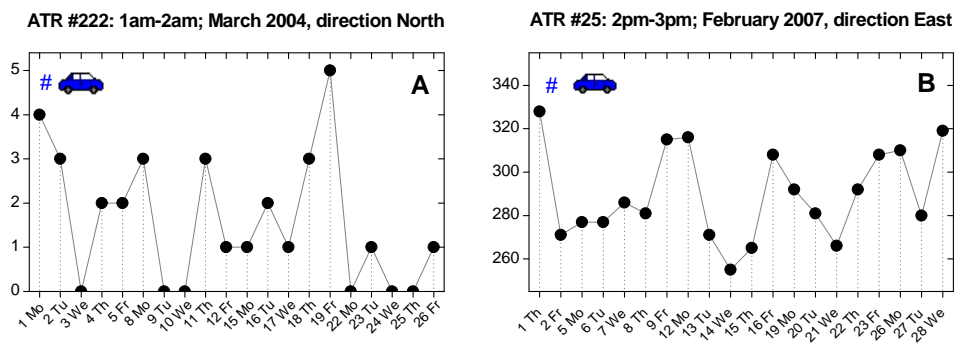,width=\columnwidth,angle=0}}
\caption{\label{series} Two examples of  time series considered in our analysis. The series represent number of cars counted by a single ATR during 20 working days in one month in the period 2002-2007. A) ATR no.~222, 1am-2am, March~2004, direction: North. B) ATR no.~25, 2pm-3pm, February 2007, direction: East.}
\end{figure}

Hourly numbers of cars passing through observation points located on interstates, trunk highways, county state-aid highways, and municipal state-aid streets at various locations throughout Minnesota were retrieved from the Minnesota Department of Transportation \cite{Minnesota}. The traffic intensity had been recorded by 72 automatic traffic recorders (ATRs) from 2002 to 2007. The datasets, from which mean and variance were calculated, include the number of cars observed by a single recorder and passing in only one direction at a given hour in all weekdays of a month (consult Figure~\ref{series} for better understanding).

\begin{figure}
\centerline{\epsfig{file=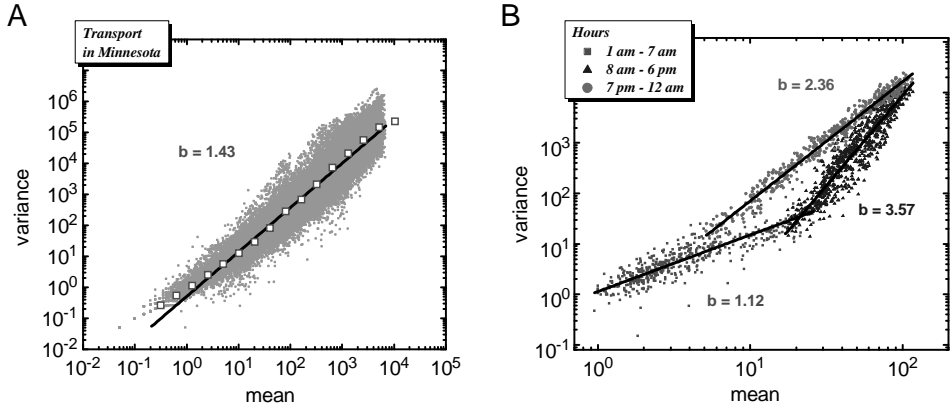,width=\columnwidth,angle=0}}
\caption{\label{real} Taylor's fluctuation scaling in transportation network of Minnesota. A) Traffic intensity as measured by all automatic traffic recorders in Minnesota in 2007. B) Daily fluctuations in traffic measured by a single recorder (ATR no. 222) in 2002-2007. In the figure, full points correspond to raw data (i.e. mean and variance calculated according to description given in the text), the open symbols express logarithmic binning of the data, and the solid lines represent their linear fits (in log-log scale).}
\end{figure}

In Figure~\ref{real}, one can see that although the whole data obey Taylor's fluctuation scaling with the characteristic exponent $b = 1.43$, in reality, the data are very heterogeneous and the parameter $b$ characterizing traffic recorded by a single ATR may change dramatically from hour to hour. For example, the car flow (i.e. the number of cars counted within a given time window) as measured by ATR no. 222 possesses three distinct phases corresponding to non-overlapping periods of time by day, i.e. night hours: 1am-7am, working hours: 8am-6pm, evening: 7pm-12am. The phases have different values of the characteristic parameter $b$. It is meaningful that during night hours the value of $b=1.12$ is very close to unity \footnote{The scaling parameter $b=1$ corresponds to Poissonian fluctuations $\langle N^2\rangle-\langle N\rangle^2=\langle N\rangle$, cf.~Eq.~(\ref{taylor}).}, which suggest motion of independent (uncorrelated) cars. During working hours traffic is characterized by the largest value of $b=3.57$. In this period, due to mutual interactions (correlations) between cars, fluctuations in traffic flow are significant and congestion (traffic jam) can be observed. Our results clearly show that in the evening, nature of traffic recorded by the considered ATR changes again. The value of $b=2.36$ attributing the new phase is lower than the corresponding value characterizing the earlier period. It implies that in the evening traffic becomes more homogeneous in comparison with working hours.

Our results show that the characteristic exponent $b$ in Taylor's power law for fluctuation may be used to characterize different phases of traffic flow. Larger values of the parameter characterize stronger collective phenomena manifesting themselves in congestion effects \footnote{It is worth to mention that a similar meaning of the parameter $b$ has been also found in other complex systems. In ecology, larger values of $b$ give evidence for stronger aggregation of individuals in different populations~\cite{1961NatureTaylor}. In genetics, they characterize stronger clustering of genes in chromosomes \cite{2004Kendal}.}. We also show that Taylor's law characterizing various complex systems may comprise of several laws with different characteristic exponents. Doing such a decomposition one can better understand the considered systems.

\section{Fluctuation scaling in the Nagel-Schreckenberg model}\label{SecNaSch}

\subsection{Description of the model}

During the last 20 years, a number of cellular automaton models have been proposed in order to better understand complex traffic phenomena and reproduce the empirical data, such as the spontaneous formation of jams \cite{2000PhysRep,2001RMPHelbing}. A particularly simple, pioneering model for single-lane traffic has been proposed by Nagel and Schreckenberg in 1992 \cite{1992NaSch}. Further in the paper we show that the model, called the NaSch model, is characterized by Taylor's fluctuation scaling with non-trivial values of the characteristic exponent $b$.

In the NaSch model a lane is represented by a one-dimensional array of $L$ sites. Each site may either be occupied by one of $N$ cars, or it may be empty ($N\leq L$). Each vehicle $i=1,2,\dots,N$ has an integer velocity $v_i$ with values between $0$ and $v_{max}$. State of the system at time $t+1$ is obtained from its state at time $t$ by applying the following rules to all cars at the same time:
\begin{enumerate}
\item[(i)] \textit{Acceleration:} unless velocity $v_i$ of a car is lower than $v_{max}$ it is advanced by one
    \begin{equation}
    v_i\rightarrow\mbox{min}[v_i+1,v_{max}].
    \end{equation}
\item[(ii)] \textit{Slowing down (due to other cars):} if the distance $d_i$ to the next car ahead is not larger than $v_i$ the speed is reduced to avoid accidents
    \begin{equation}
    v_i\rightarrow\mbox{min}[v_i,d_i].
    \end{equation}
\item[(iii)] \textit{Randomization:} with probability $p$ the velocity of a car (if greater than zero) is decreased by one
    \begin{equation}
    v_i\rightarrow v_i-1\mbox{   with probability $p$}.
    \end{equation}
\item[(iv)] \textit{Car motion (update of positions):} each car is advanced $v_i$ sites.
\end{enumerate}

\subsection{Simulation procedure}

Numerical simulations of the model have been performed on one-dimensional lattice of size $L=10^5$ with periodic boundary conditions. Each simulation has started at random initial conditions, i.e. random localization of $N$ cars on a lane (further in the text we rather use the parameter $\rho=N/L$ than $N$) and random initial velocities. Then, we have updated state of the system in accordance with the rules described above. Collection of data has started after the first $L$ time steps.

\subsection{Taylor's law in the model}

\begin{figure}
\centerline{\epsfig{file=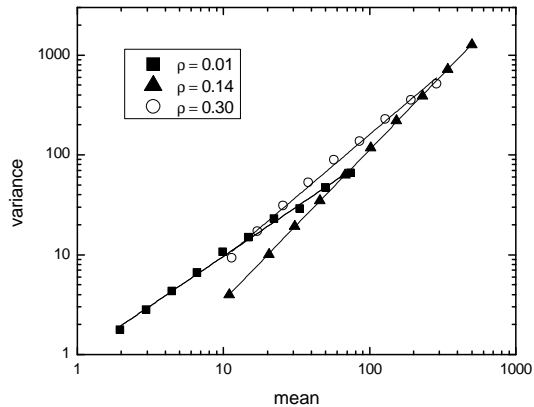,width=7cm,angle=0}}
\caption{\label{NaSchT} Fluctuation scaling in time series representing car motion in the NaSch model described by the parameters $v_{max}=10$, $p=0.1$ and different values of the car density $\rho$.}
\end{figure}

In our analysis we have concentrated on time series whose data points represent the number of cars passing through arbitrary lattice site in a given time period $\tau$ (cf.~Figure~\ref{series}). Having such a time series one can calculate its mean and variance (both depending on $\tau$). Doing so for different window sizes $\tau=10,20,50,100 \dots$ and plotting the obtained mean-variance points in a log-log graph we have obtained Taylor's power laws characterizing various stages of the traffic flow described by the set of model parameters $v_{max}$, $p$ and $\rho$. In Figure~\ref{NaSchT} we have shown three scaling laws describing car motion in the NaSch model with $v_{max}=10$, $p=0.1$ and different values of the car density $\rho$.

\begin{figure}
\centerline{\epsfig{file=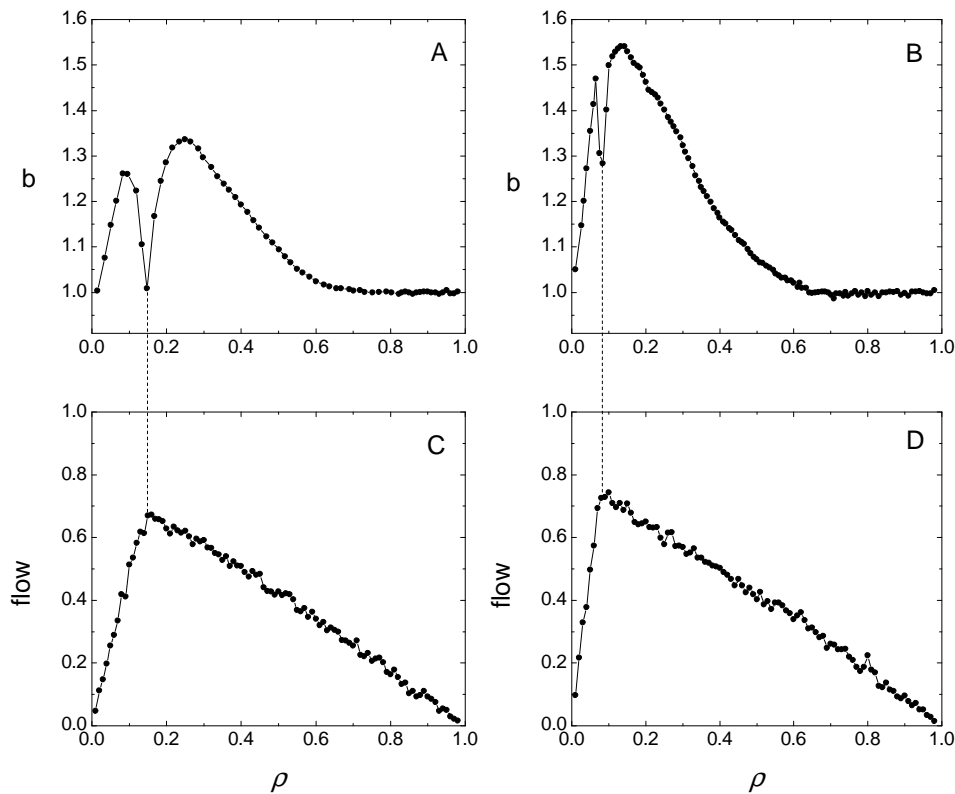,width=\columnwidth,angle=0}}
\caption{\label{NaSchb} Scaling exponent $b$ as a function of the car density $\rho$ in the NaSch model characterized by: A) $v_{max}=5$ and $p=0.1$, B) $v_{max}=10$ and $p=0.1$. Fundamental diagrams describing efficiency of the car flow in the two considered cases: C) $v_{max}=5$ and $p=0.1$, D) $v_{max}=10$ and $p=0.1$.}
\end{figure}

In Figure~\ref{NaSchb}A and B one can see how the scaling parameter $b$ characterizing fluctuations in the considered model depends on the car density $\rho$ for two given values of the speed limit $v_{max}=5$ and $10$ and the fixed probability of slowing down $p=0.1$. Our simulations show that in the vicinity of the transition from the laminar traffic flow to the congested phase, it means in the vicinity of the maximum in the fundamental diagram (see Figure~\ref{NaSchb}C and D), values of the parameter $b$ are significantly larger than outside the region. It means that the transition is turbulent. It is accompanied by large fluctuations, just like in continuous phase transitions. The sudden reduction of the scaling exponent $b$ exactly at the critical density (dashed lines in the considered figure) is due to finite size effects, which cause that in the vicinity of the transition point traffic becomes locally regular.

The result are consistent with our previous observations obtained for real traffic, according to which below the transition point the growing density of cars causes higher values of $b$. These in turn show evidence of strongly inhomogeneous, turbulent traffic.

\section{Concluding remarks}\label{SecSum}

A short summary of the paper has been given in the Introduction (Section~\ref{intro}), therefore we avoid it here. Nevertheless, we would like to draw the readers attention to issues related to the origin of the Taylor's law for fluctuation scaling. At present, universality of the law is widely recognized. The list of scientific disciplines in which Taylor's fluctuation scaling has been observed encompasses: genetics, epidemiology, medicine, physics, economy, computer and social sciences. It allows one to think, that there must exist a common mechanism underlying this law. In our recent paper \cite{2009Fronczak} we show that the conjecture is justified. In the paper, for the first time we give a universal, microscopic explanation of the law. Supported by real world observations ranging from insect and bird populations, through the human chromosome to traffic intensity in transportation network, we show that the law results from the density of state function characterizing the considered systems (the concept borrowed from statistical physics).

\section{Acknowledgments}

The paper was supported by the Polish Ministry of Science, grant no. 496/N-COST/2009/0.

\end{document}